\documentclass[twocolumn, nofootinbib]{revtex4}
\usepackage[utf8]{inputenc}
\usepackage{amsmath}
\usepackage{graphicx}
\usepackage{float}
\usepackage{tablefootnote}
\restylefloat{table}
\usepackage[T1]{fontenc}
\usepackage{mathtools}  
\usepackage{amssymb}
\usepackage{color}
\usepackage{cancel}
\usepackage{framed}
\usepackage{comment}
\usepackage{mathtools}
\usepackage{hyperref}
\usepackage{chngcntr}
\usepackage[normalem]{ulem}
\usepackage[flushleft]{threeparttable}
\usepackage{multirow}
\usepackage[dvipsnames]{xcolor}

\counterwithin{equation}{section}

\newcommand{\nn}{\nonumber}

\begin{document}


\title{\Large\bfseries Interacting dark energy axions in light of the Hubble tension }

\author{Ennis Mawas}
\email{mawases@mail.uc.edu}
\affiliation{Department of Physics, University of Cincinnati, Cincinnati, OH 45221 USA}
\author{Lauren Street}
\email{streetlg@mail.uc.edu}
\affiliation{Department of Physics, University of Cincinnati, Cincinnati, OH 45221 USA}
\author{Richard Gass}
\email{gassrg@ucmail.uc.edu}
\affiliation{Department of Physics, University of Cincinnati, Cincinnati, OH 45221 USA}
\author{L.C.R. Wijewardhana}
\email{rohana.wijewardhana@gmail.com}
\affiliation{Department of Physics, University of Cincinnati, Cincinnati, OH 45221 USA}

\begin{abstract}
A current problem within the $\Lambda$CDM framework is the tension between late and early time measurements of the Hubble parameter today, $H_0$.  We entertain the possibility that dark energy modeled as multiple interacting axion-like-particle species can alleviate the current Hubble tension.  We then test these parameters against the milder tension between the CMB and large scale structure (LSS) observations of $\sigma_{8}$ to ensure that these models do not exacerbate the tension.  We find that there exist parameter spaces for models of two and three axion-like-particles which can potentially alleviate the Hubble tension as well as the $\sigma_8$ tension.
\end{abstract}

\maketitle

\section{Introduction}
\label{intro}

The $\Lambda$CDM framework is the most widely accepted model that encapsulates dark energy, matter, and radiation. The $\Lambda$CDM model fits well with a wide scope of observations from the cosmic microwave background (CMB), the accelarated expansion of the universe, and general relativity \cite{Condon:2018eqx}. However, motivation to look for a new model stems from the newfound theoretical and observational disagreements with $\Lambda$CDM. In this work, we shall focus on the recent observations from the CMB that are in tension with the data obtained from other cosmological investigations if given a $\Lambda$CDM universe. Explicitly, the $H_0$ tension refers to the $4\sigma$ to $6\sigma$ discrepancy between measurements of $H_0$ using CMB anisotropy data from Planck  ($H_0 = 67.36 \pm 0.54 \, \text{km}/\text{s}/\text{Mpc}$ \cite{2020}) and late-time (local) measurements from the SH0ES collaboration ($H_0$ = $73.3 \pm 0.8 \, \text{km}/\text{s}/\text{Mpc}$ \cite{Riess:2020fzl}). There have been numerous families of models proposed that extend or change the $\Lambda$CDM model with the intent of alleviating the discrepancy between the two measurements. Examples of such models include early-time solutions \cite{Poulin:2018cxd,Lin:2019qug,Hill:2020osr,Sakstein:2019fmf,Niedermann:2020dwg}, interacting dark energy and dark matter solutions \cite{DiValentino:2017iww,Yang:2018uae,DiValentino:2019ffd,Yang:2018euj,Yang:2019uzo,Wang:2016lxa}, and late-time solutions \cite{Aghanim:2018eyx,DiValentino:2019dzu,Yang:2020zuk,Cai:2021wgv,Wang:2016och}.  

Here, we analyze a potential late-time solution to the Hubble tension, which involves changing the expansion history after matter-radiation decoupling by making use of an evolving dark energy equation of state $\omega(z)= p(z)/\rho(z)$. A common problem with late-time solutions is that they may solve the $H_0$ tension, but they tend to have trouble agreeing with baryonic acoustic oscillations (BAO) and supernova Type Ia (Pantheon) data \cite{DiValentino:2021izs,Knox:2019rjx,Arendse:2019hev}. Early-time solutions that solve the Hubble tension, by reducing the sound horizon, agree with BAO data but tend to fall into tension with late-time local measurements \cite{Arendse:2019hev}. As we shall discuss in more detail in a later section, it has been suggested that it can still be worthwhile to analyze late-time solutions despite the possible tensions that may arise \cite{Arendse:2019hev,Jedamzik:2020zmd,Beenakker:2021vff,Krishnan:2020obg}. Therefore, we propose a model that demonstrates a set of parameters that may simultaneously alleviate the $H_0$ and $\sigma_8$ tensions. A best-fit parameter analysis with observational constraints from cosmological data sets is to be done in future work.

The model discussed in this paper is based off of a multiple interacting axion-like-particle (ALP) model formulated by D'Amico et al. \cite{DAmico:2016jbm}. ALPs are of interest because they have been suggested as candidates of dark matter \cite{Arias:2012az,Visinelli:2017imh} as well as candidates of dark energy \cite{Hlozek:2014lca,Visinelli:2018utg}. They have also been suggested to alleviate the Hubble tension, among other tensions, when coupled to the Higgs field \cite{Fung:2021fcj,Fung:2021wbz}. In this work, we extend the D'Amico et al. model \cite{DAmico:2016jbm} to include three interacting dark energy ALPs and then search for parameter spaces in the two and three interacting models that alleviate the $H_0$ tension. We show that there exist parameters in this model such that the Hubble parameter, $H(z)$, normalized to the Planck measurement \cite{2020} at $z = 1100$, evolves to $H_0$ $\approx$ $74 \, \text{km}/\text{s}/\text{Mpc}$ at $z = 0$ \cite{Riess_2011,Riess_2016,Riess_2019,Riess:2020fzl}. We then ensure that the parameters that solve the $H_0$ tension do not exacerbate the $\sigma_8$ tension. The $\sigma_8$ tension is the discrepancy between the CMB and large scale structure (LSS) measurements of the amplitude of the linear power spectrum, $\sigma_{8}$, on the scale of $8 h^{-1} \text{Mpc}$ \cite{Macaulay:2013swa,Ade:2015xua,2020,Heymans:2020gsg,Abbott:2017wau}. We use the method formulated by Barros et al. \cite{Barros:2018efl} for approximating the value of $\sigma_8$ for coupled quintessence models, to predict the values of $\sigma_{8}$ in the ALP models.

In Sec. \ref{sec:formalism}, we outline the general formalism describing $N$ interacting ALPs and their evolution, as well as the evolution of the Hubble parameter.  We then explore scenarios corresponding to $N=2$ and $N=3$.  In Sec. \ref{sec:Hubble}, we discuss possible parameter spaces for two and three interacting ALP models that can alleviate the Hubble tension.  In Sec. \ref{sec:perturbs}, we outline the formalism describing matter perturbations in this model and show that parameter spaces that alleviate the Hubble tension can also alleviate the $\sigma_8$ tension.  In Sec. \ref{Sec: sound horizon}, we describe a common difficulty that models such as the one described here run into, namely the sound horizon problem.  Finally, we conclude in Sec. \ref{sec:conclusion}.

\section{Formalism}\label{sec:formalism}
Generalizing the approach of D'Amico et al. \cite{DAmico:2016jbm}, we consider a number of interacting ALPs through the self-interaction potential, 
\begin{align}\label{eq:int_pot}
V_{\text{ALP}}\left(\vec{\phi}\right) &=\sum_{i=1}^N \mu_i^4 \left[1 - \cos\left(\frac{\phi_i}{f_i}\right)\right] 
\nn\\
&+
\sum_{i=1}^{N-1} \sum_{j=i+1}^{N} \mu_{i,j}^4
 \left[1 - \cos\left(\frac{\phi_i}{f_i} - n_{i,j} \frac{\phi_j}{f_j}\right)\right],
\end{align}
where each $\phi_i$ is a real scalar field, $f_i$ is the decay parameter, and $\mu_i$ is the self-interaction constant of the $i$-th ALP.  Here, the constants parameterizing the interaction between the $i$-th and $j$-th ALPs are the interaction constant $\mu_{i,j}$ and the mixing constant  $n_{i,j}$.

We now consider how multiple species of ALPs affect the evolution of the Hubble parameter.  Assuming some fractional energy density of matter $\Omega_{m,0}$ and radiation $\Omega_{r,0}$ today and zero curvature, the Friedmann and Klein-Gordon (KG) equations are,
\begin{align}\label{eq:klein-gordon}
H^2 &= H_0^2 \left[\frac{\rho_\text{ALP}}{3 H_0^2 M_P^2} + \frac{\Omega_{m,0}}{a^3} + \frac{\Omega_{r,0}}{a^4}  + \Omega_\Lambda \right],
\nn \\
0&=
\frac{d^2 \phi_i}{d t^2} + 3 H \frac{d \phi_i}{d t} + \frac{d \, V_\text{ALP}}{d \phi_i},
\end{align}
where $H_0$ is the Hubble parameter today, $M_P = 1/\sqrt{8 \pi G}\approx 2.4 \times 10^{18}\, \text{GeV}$, and
\begin{align}
\rho_\text{ALP} = \frac{1}{2} \left[ \left(\frac{d \phi_1}{d t}\right)^2 + ... + \left(\frac{d \phi_N}{d t}\right)^2 \right] + V_\text{ALP}\left(\vec{\phi}\right).
\end{align}
Scaling parameters as,
\begin{align}\label{eq:scaling}
\begin{matrix}
\tilde{\mu} \equiv \frac{\mu_{\text{i,j}}}{(H_0^2 M_P^2) }& \qquad \tilde{f} \equiv \frac{f_{\text{i}}}{M_P} & \qquad \tilde{V}_\text{ALP} \equiv \frac{V_\text{ALP}}{(H_0^2 M_P^2) } 
\\
\\
x_i \equiv \frac{\phi_i }{f_i }& \qquad \tau \equiv H_0 t & \qquad \tilde{H} \equiv \frac{H}{H_0}
\end{matrix},
\end{align}
the Friedmann and KG equations become,
\begin{align}
\label{eq:FE}
\tilde{H}^2  &= \frac{1}{3} \left[\frac{1}{2} \sum_{i=1}^N \tilde{f}_i^2 \, \left(\frac{d x_i}{d \tau}\right)^2 + \tilde{V}_\text{ALP}\left(\vec{x}\right)\right]\nn\\ 
&+ \frac{\Omega_{m,0}}{a^3} + \frac{\Omega_{r,0}}{a^4} + \Omega_\Lambda,
\nn \\
0&=
\tilde{f}_i \, \frac{d^2x_i}{d\tau^2} + 3 \tilde{H} \tilde{f}_i \, \frac{d x_i}{d \tau}   + \tilde{f}_i^{-1}\frac{d \, \tilde{V}_\text{ALP}}{d x_i}.
\end{align}
In the next sections, we outline the formalism for the examples of two interacting ALPs, which has been analyzed by D'Amico et al. \cite{DAmico:2016jbm}, and three interacting ALPs.

\subsection{Two and Three Interacting Models}
Consider now the example of two interacting ALPs as in D' Amico et al. \cite{DAmico:2016jbm}.  In this case, the interaction potential (Eq. (\ref{eq:int_pot})) becomes,
\begin{align}\label{eq:int_pot_2ALPs}
\tilde{V}_{\text{ALP}}\left(\vec{x}\right) &=\tilde{\mu}_1^4 \left[1 - \cos\left(x_1\right)\right] +  \tilde{\mu}_2^4 \left[1 - \cos\left(x_2\right)\right]
\nn\\
&+ \tilde{\mu}_{1,2}^4 \left[1 - \cos\left(x_1 - n_{1,2} x_2\right)\right].
\end{align}
The Friedmann and KG equations (Eq. (\ref{eq:FE})) for this model are,
\begin{widetext}
\begin{align}
\label{eq:FE_2ALPs}
\tilde{H}^2 &= \frac{1}{3} \left[\frac{1}{2} \sum_{i=1}^2 \tilde{f}_i^2\, \left(\frac{d x_i}{d\tau}\right)^2 
+ \tilde{V}_{\text{ALP}}\left(\vec{x}\right)\right]
+ \frac{\Omega_{m,0}}{a^3} + \frac{\Omega_{r,0}}{a^4} + \Omega_\Lambda,
\nn \\
0&=
\tilde{f}_1 \, \frac{d^2 x_1}{d\tau^2}+ \, 3 \tilde{H} \tilde{f}_1\, \frac{dx_1}{d\tau}   +  \frac{\tilde{\mu}_1^4}{\tilde{f}_1}\sin\left(x_1\right)
+ \frac{\tilde{\mu}_1^4}{\tilde{f}_1} \sin \left(x_1 - n_{1,2} x_2 \right),
\nn\\
0&=\tilde{f}_2 \, \frac{d^2 x_2}{d\tau^2}+  3 \tilde{H} \tilde{f}_2\, \frac{d x_2}{d\tau} 
 +  \frac{\tilde{\mu}_2^4}{\tilde{f}_2} \sin\left(x_2\right)
- n_{1,2}\, \frac{\tilde{\mu}_{1,2}^4}{\tilde{f}_2} \sin \left(x_1 - n_{1,2} x_2 \right).
\end{align}
\end{widetext}
Similarly, for three ALPs one can take $N = 3$ and,  for simplicity,  take $\mu_{1,3} = 0$ in Eq. (\ref{eq:int_pot}) to obtain,
\begin{align}\label{eq:int_pot_3ALPs}
\tilde{V}_{\text{ALP}}\left(\vec{x}\right)& =\sum_{i=1}^3 \tilde{\mu}_i^4 \left[1 - \cos\left(x_i\right)\right]   
\nn\\
&+ \tilde{\mu}_{1,2}^4 \left[1 - \cos\left(x_1 - n_{1,2} x_2\right)\right] 
\nn\\
&+ \tilde{\mu}_{2,3}^4 \left[1 - \cos\left(x_2 - n_{2,3} x_3\right)\right],
\end{align}
and the Friedmann and KG equations (Eq. (\ref{eq:FE})) for the three ALP model are,
\begin{widetext}
\begin{align}\label{eq:FE_3ALPs}
\tilde{H}^2 &= \frac{1}{3} \left[\frac{1}{2} \sum_{i=1}^3 \tilde{f}_i^2\, \left(\frac{d x_i}{d \tau}\right)^2  + \tilde{V}_{\text{ALP}}\left(\vec{x}\right)\right]+ \frac{\Omega_{m,0}}{a^3} + \frac{\Omega_{r,0}}{a^4} + \Omega_\Lambda,
\nn\\
0&=\tilde{f}_1 \, \frac{d^2 x_1}{d\tau^2} + 3 \tilde{H} \tilde{f}_1\, \frac{d x_1}{d\tau}   + \frac{\tilde{\mu}_1^4}{\tilde{f}_1}\sin\left(x_1\right) 
+ \frac{\tilde{\mu}_{1,2}^4}{\tilde{f}_1}\sin\left(x_1 - n_{1,2} x_2 \right),
\nn\\
0&=\tilde{f}_2 \, \frac{d^2 x_2}{d\tau^2}  + 3 \tilde{H} \tilde{f}_2\, \frac{d x_2}{d \tau} + \frac{\tilde{\mu}_{2}^4}{\tilde{f}_2}\sin\left(x_2\right)  - n_{1,2}\,\frac{\tilde{\mu}_{1,2}^4}{\tilde{f}_2} \sin \left(x_1 - n_{1,2} x_2 \right)
+  \frac{\tilde{\mu}_{2,3}^4}{\tilde{f}_2} \sin \left(x_2 - n_{2,3} x_3 \right),
\nn\\
0&=\tilde{f}_3 \, \frac{d^2x_3}{d\tau^2} + 3 \tilde{H} \tilde{f}_3\, \frac{dx_3}{d\tau} +  \tilde{f}_3^{-1}\left[\tilde{\mu}_3^4 \sin\left(x_3\right) - n_{2,3}\, \tilde{\mu}_{2,3}^4 \sin \left(x_2 - n_{2,3} x_3 \right)\right].
\end{align}
\end{widetext}

As in \cite{DAmico:2016jbm}, we choose parameters and initial conditions for the $x_N$ fields to ensure that the fields do not cross over a minimum of the potential (Eq. (\ref{eq:int_pot})). If the potential reaches a minimum, the energy density of the axions becomes negligible and we cannot reproduce dark energy. Hence, we choose initial conditions such that all of the energy density due to dark energy is made up of the $x_N$ fields.  By choosing parameters that do not cross over a minimum of the potential, one can control this model to yield the energy-density needed to fit the necessary constraints from cosmological data.  One can imagine the numerous possibilities of the parameter spaces that may alleviate the $H_0$ and $\sigma_8$ tensions. Therefore, it is important to do a simultaneous fit to the available cosmological data to unveil the preferred parameters. In this work, we discuss examples of parameters that alleviate these tensions as a proof of concept.

\begin{figure}
	\centering
	\includegraphics[width=0.37\textwidth]{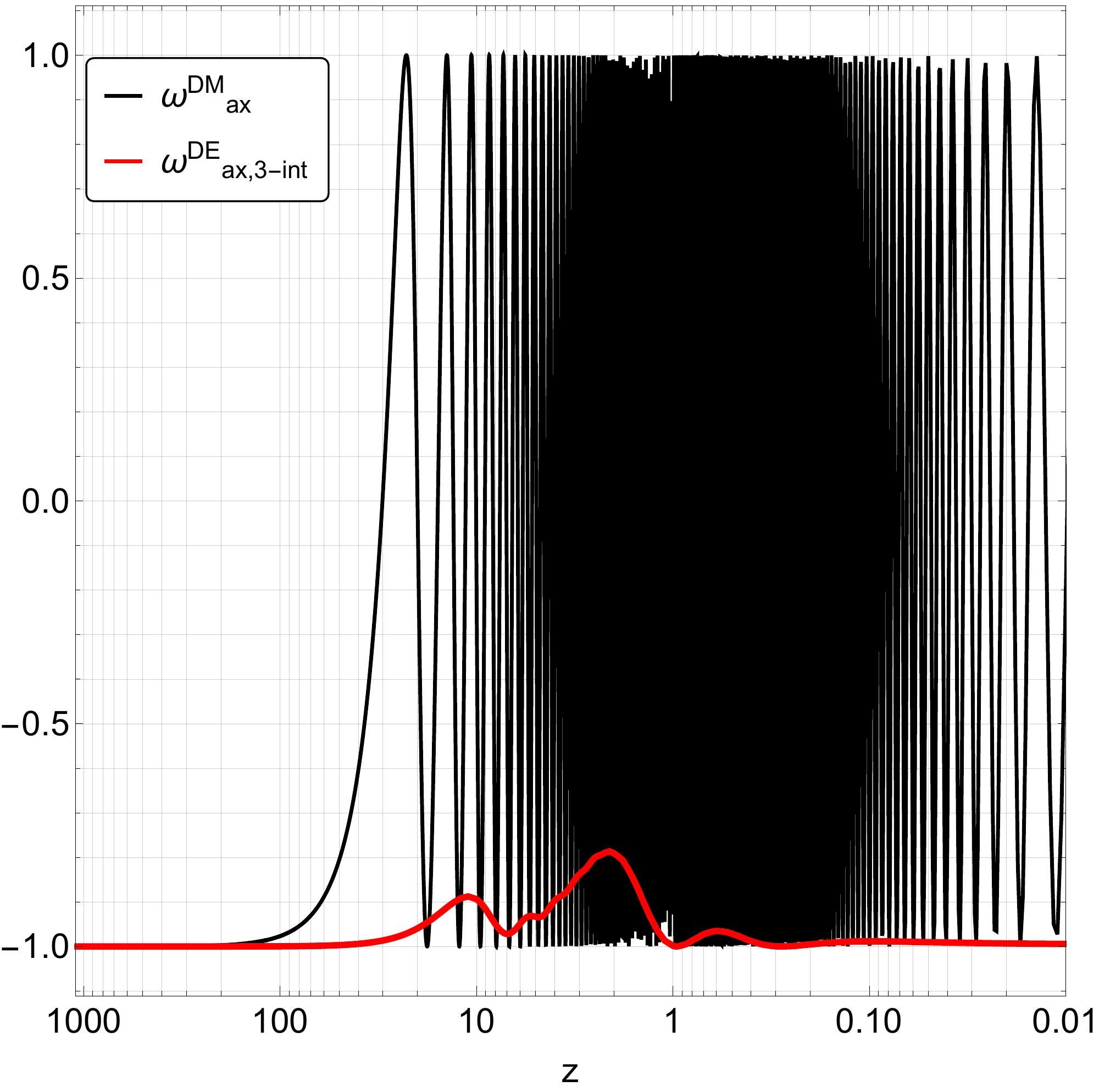}
	\caption{Equation of state of a non-interacting dark matter axion compared to the equation of state of three interacting dark energy axions with parameters given in Table \ref{tab:ALP_params}. The subscript `ax` refers to quantities in the ALP models. }
	\label{fig:wplot2}
\end{figure}

Choosing parameters and initial conditions that ensure the $x_N$ fields are still rolling at late times allows the ALPs to make up all of the dark energy density.  Hence, we set $\Omega_\Lambda = 0$ in Eq. (\ref{eq:FE_2ALPs}). As shown in Figs. \ref{fig:wplot2} and \ref{fig:w}, we find that the equation of state settles to $w(z) = -1$ at late times, reinforcing the claim that these ALPs act like dark energy. 

 It is interesting to compare this phenomenon to the usual cosmological evolution of an ALP which does not interact with other ALPs through Eq. (\ref{eq:int_pot}) and evolves like dark matter.  To simulate a dark matter axion, we choose parameters that ensure that the mass of the axion, $m_a$,  is $m_a > H_0$, where $m_a^2 = \frac{\mu_1^4}{f_1^2}$. We use parameters $\mu_1^4 = 10 H_0^2 M_p^2$ and $f_1 = 0.01 M_p$. As shown in Fig. \ref{fig:wplot2}, once the Hubble parameter becomes of the order of the mass of the ALP its equation of state begins to oscillate rapidly between $w(z) = -1$ and $w(z) = 1$, in effect averaging out to $w(z) = 0$. In this case, the ALP energy density scales as $\rho \sim a^{-3}$ and can be treated as matter \cite{Marsh:2015xka}. For the model discussed here, however, we add the entire dark matter energy density in by hand  as $\frac{\Omega_{m,0}}{a^3}$. We note that one could modify these interacting models to include one or more interacting species of ALPs that contribute to the dark matter energy density.  These ALPs would have begun contributing to the matter energy density long before last scattering.  For this work, we begin our evolution after last scattering. Hence by this time their density can be put into the equations of motion by hand.

\section{The Hubble Tension}\label{sec:Hubble}
We now consider ALP parameters which can potentially alleviate the Hubble tension.  We set the cosmological parameters to those observed from Planck \cite{2020} and solve the Friedmann and KG equations within the redshift range $z = [1100,0]$.  In order that the models discussed alleviate the Hubble tension, we must find that the ALP parameters chosen result in an increase of the Hubble parameter at $z = 0$ to that observed by late-time local measurements such as the SH0ES collaboration \cite{Riess_2011,Riess_2016,Riess_2019,Riess:2020fzl}.  

\begin{figure}[H]
	\centering
	\includegraphics[width=0.37\textwidth]{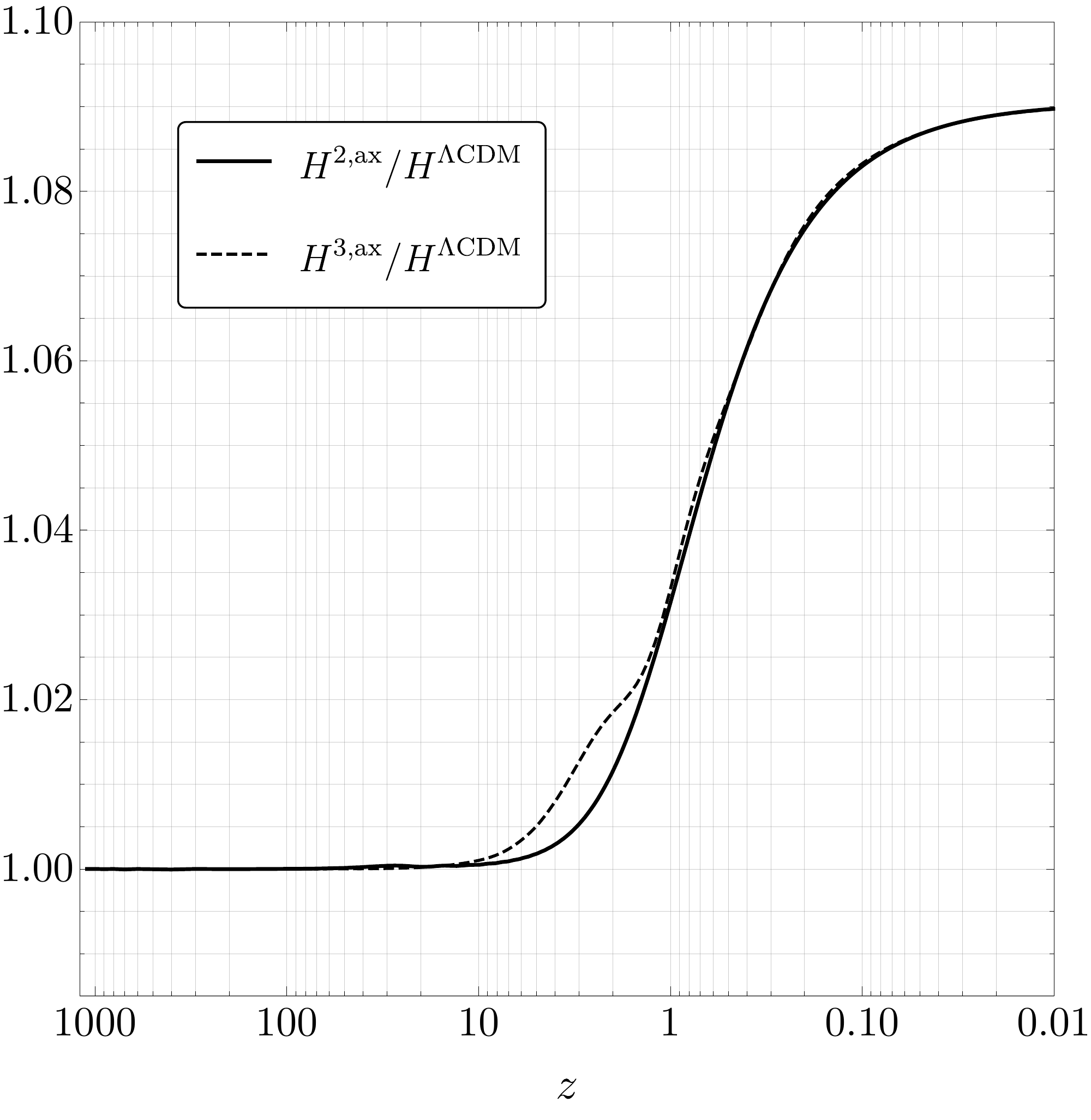}
	\,
	\includegraphics[width=0.37\textwidth]{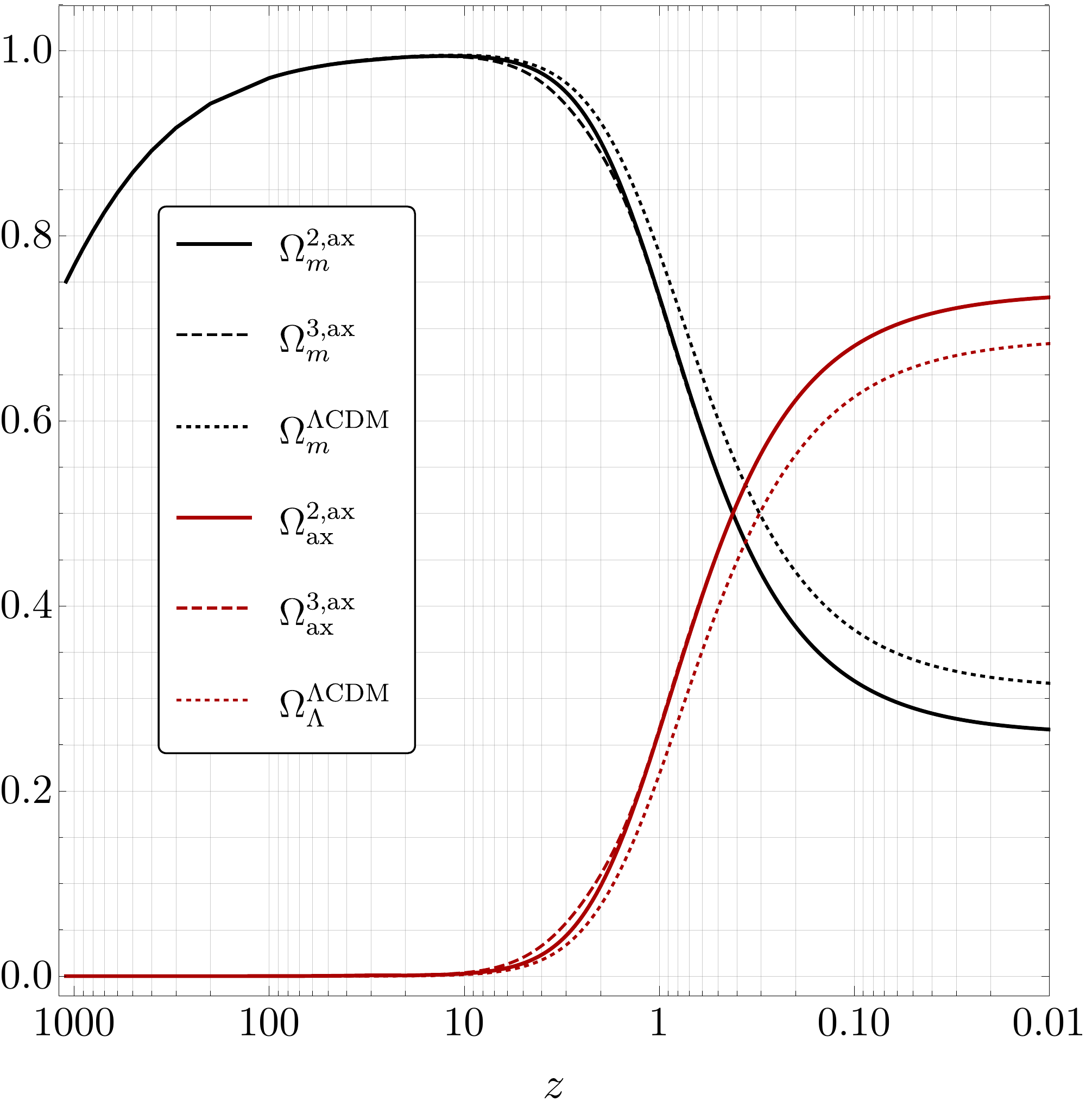}
	\caption{Results for the two and three interacting ALP models with parameters given in Table \ref{tab:ALP_params}.  \textbf{Top:}  Ratio of the resulting Hubble parameter to the Hubble parameter in the $\Lambda$CDM model.  \textbf{Bottom:} Fractional energy densities $\Omega_{m}$ and $\Omega_{\text{ax}}$ for the interacting models and $\Omega_{m}$ and $\Omega_\Lambda$ for the $\Lambda$CDM model.}
	\label{fig:ratio_dens}
\end{figure}

For the two ALP model we solve Eq. (\ref{eq:int_pot_2ALPs}) with Eq. (\ref{eq:FE_2ALPs}), while for the three ALP model we solve Eq. (\ref{eq:int_pot_3ALPs}) with Eq. (\ref{eq:FE_3ALPs}). The parameters used are shown in Table \ref{tab:ALP_params}.  We begin the numerical integration of the equations of motion around the time of last scattering ($z\approx 1100$) assuming parameters as observed by Planck  \cite{2020}.  The top panel of Fig. \ref{fig:ratio_dens} shows the ratio of the Hubble parameter in the two and three ALP models to that in the $\Lambda$CDM model with the initial conditions normalized to the Planck data \cite{2020}.  In both the two and three interacting ALP models, we estimate the Hubble parameter today as $H_0^\text{int} \approx 1.09 \, H_0^{\Lambda\text{CDM}} \approx 74 \, \text{km}/\text{s}/\text{Mpc}$.  Therefore, the tension in the observed values of $H_0 \approx 68 \, \text{km}/\text{s}/\text{Mpc}$ as observed by Planck  \cite{2020} and $H_0 \approx 74 \, \text{km}/\text{s}/\text{Mpc}$ as observed by the SH0ES collaboration \cite{Riess_2011,Riess_2016,Riess_2019,Riess:2020fzl} can potentially be alleviated by the models discussed in this work.

\begin{widetext}
\begin{table*}
	\caption{\label{tab:ALP_params}Parameters for interacting ALPs.  Each $\mu_i^4$ and $\mu_{i,j}^4$ is in units of $H_0^2 M_P^2$, while each $f_i$ and $\phi_{i,\text{in}}$ is in units of $M_P$.}
	\begin{ruledtabular}
		\begin{tabular}{cccccccccccccc}
			& $\mu_1^4$	
			& $\mu_2^4$	
			& $\mu_3^4$	
			& $\mu_{1,2}^4$	
			& $\mu_{2,3}^4$	
			& $n_{1,2}$
			& $n_{2,3}$
			& $f_1$	
			& $f_2$	
			& $f_3$	
			& $\phi_{1, \text{in}}$	
			& $\phi_{2, \text{in}}$	
			& $\phi_{3, \text{in}}$	
			\\ \hline
			Two Interacting ALPs 
			& $8$ 
			& $4.81$	
			& $-$
			& $30$
			& $-$
			& $40$	
			& $-$
			& $0.90$	
			& $0.68$	
			& $-$
			& $0.10$	
			& $0.78$
			& $-$	
			\\
			Three Interacting ALPs 
			& $5$ 
			& $0.72$	
			& $1$
			& $1$
			& $1$
			& $9$	
			& $12$
			& $0.20$	
			& $0.10$	
			& $0.10$
			& $0.17$	
			& $0.90$
			& $0.8$	
		\end{tabular}
	\end{ruledtabular}
\end{table*}
\end{widetext}

\begin{figure}[H]
	\centering
	\includegraphics[width=0.37\textwidth]{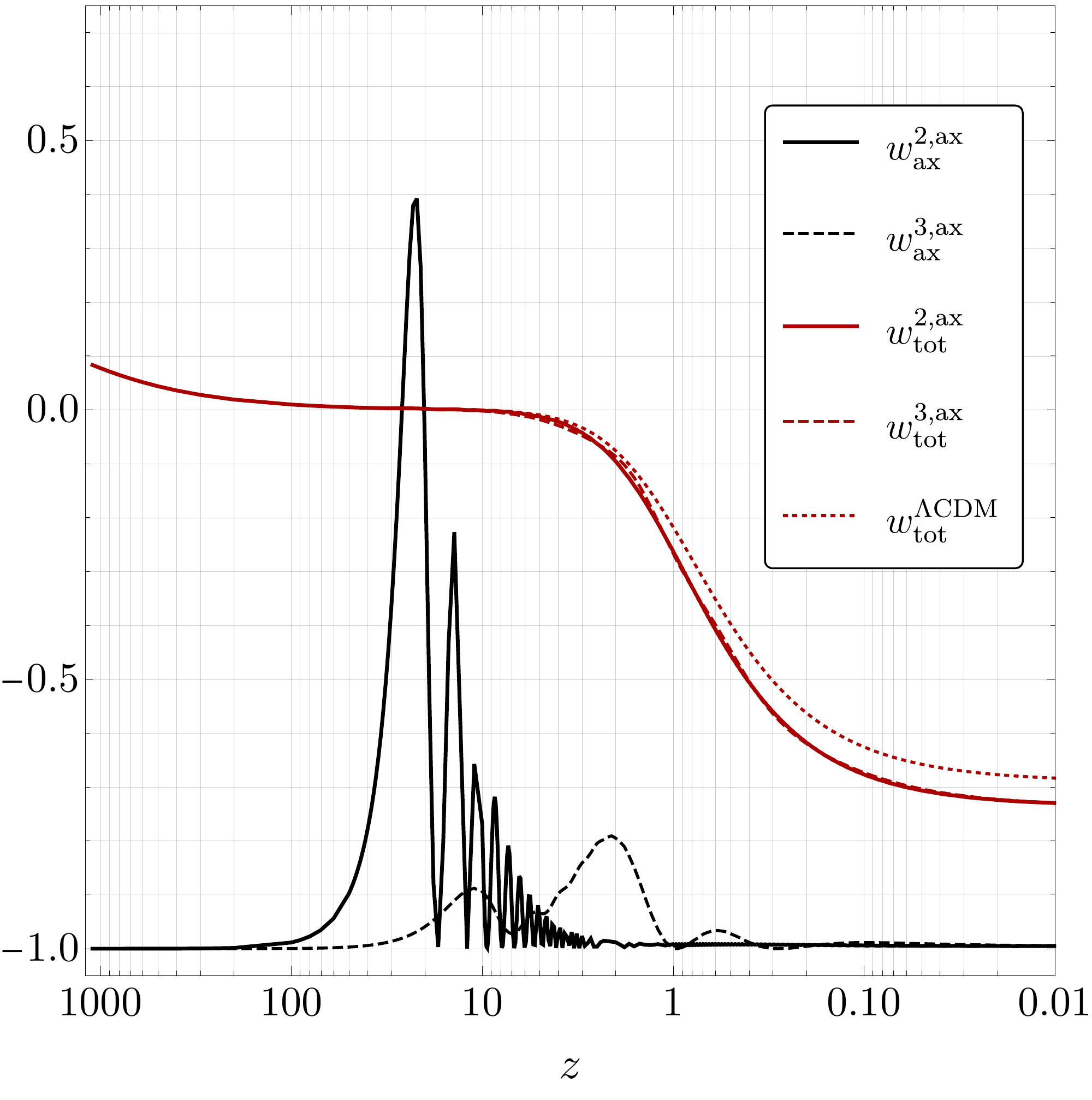}
	\caption{Axion equations of state for the two and three interacting ALP models and the total equation of state for the interacting and $\Lambda$CDM models with parameters given in Table \ref{tab:ALP_params}.}
	\label{fig:w}
\end{figure}

The bottom pannel of Fig. \ref{fig:ratio_dens} shows the results for the total fractional energy densities of dark matter and dark energy compared to the $\Lambda$CDM model.  Notice from Fig. \ref{fig:w} that the total equations of state in the interacting models decrease relative to the total equation of state in the $\Lambda$CDM model at late times.  This is presumably due to the increase in the Hubble parameter in the interacting model relative to the $\Lambda$CDM model.  Finally, notice in the top panel Fig. \ref{fig:ratio_dens} that, aside from $1 \lesssim z \lesssim 10$, the two ALP models discussed here are indistinguishable from each other when comparing the ratios of Hubble parameters or the densities.  However, as shown in Fig. \ref{fig:w}, each model is quite distinguishable when comparing the axion equations of state.  In this case, it seems that there are interacting ALP models which can potentially alleviate the Hubble tension and can also give a unique equation of state \cite{DAmico:2016jbm}.

\section{Matter Perturbations}\label{sec:perturbs}
Now, we consider how these models affects the matter perturbations, and hence the current $\sigma_8$ tension \cite{Macaulay:2013swa,Hildebrandt:2016iqg,Ade:2015xua,2020,Abbott:2017wau}.  The perturbation equations of motion are given by \cite{DAmico:2016jbm},
\begin{align}
0&=-2 \frac{d^2\delta_m}{dt^2} - 4 H \frac{d \delta_m}{dt} + 3 H^2 \Omega_m \delta_m 
\nn \\
&+ \frac{8 \pi}{M_P^2} \sum_i \left(4 \frac{d\phi_i}{dt} \frac{d \delta\phi_i}{dt} - 2 \frac{\partial V_\text{ALP}}{\partial \phi_i} \delta\phi_i\right)
\nonumber \\
0&=\frac{d^2 d\phi_i}{dt^2}+ 3 H \frac{d\delta\phi_i}{dt} + \frac{k^2}{a^2} \delta \phi_i + \sum_j \delta\phi_j \frac{\partial^2 V_\text{ALP}}{\partial \phi_i \partial \phi_j} 
\nn\\
&- \frac{d \delta_m}{dt} \frac{d \phi_i}{dt}
\end{align}
In scaled variables (Eq. (\ref{eq:scaling})),
\begin{align}
0&=-2 \frac{d^2\delta_m}{d\tau^2} - 4 \tilde{H} \frac{d\delta_m}{d\tau} + 3 \tilde{H}^2 \Omega_m \delta_m 
\nn\\
&+ 2 \sum_i \left(2 \tilde{f}_i^2 \frac{dx_i}{d\tau} \frac{d \delta x_i}{d\tau} - \frac{\partial \tilde{V}_\text{ALP}}{\partial x_i} \delta x_i\right)
\nonumber \\
0&=\tilde{f}_i \frac{d^2 \delta \phi_i}{d\tau^2} + 3 \tilde{f}_i \tilde{H} \frac{d x_i}{d\tau} + \tilde{f}_i \frac{\tilde{k}^2}{a^2} \delta x_i + \sum_j \tilde{f}_i^{-1} x_j \frac{\partial^2 \tilde{V}_\text{ALP}}{\partial x_i \partial x_j}
\nn\\
& - \tilde{f}_i \frac{ d\delta_m}{d\tau} \frac{dx_i}{d\tau}
\end{align}
where,
\begin{align}
\begin{matrix}
\tilde{k} \equiv \frac{k}{H_0} \,&  \text{and}  \,&  \delta_m \equiv P(k)k^3.
\end{matrix}
\end{align}

Where $P(k)$ is the matter power spectra. We simulate $P(k) $ using the Code for Anisotropies in the Microwave Background (CAMB) \cite{Lewis:1999bs}, where $k$ is in units of $\text{Mpc}^{-1}$ and $P(k) $ is in units of $\text{Mpc}^3$.

\begin{figure}[H]
	\includegraphics[width=0.37\textwidth]{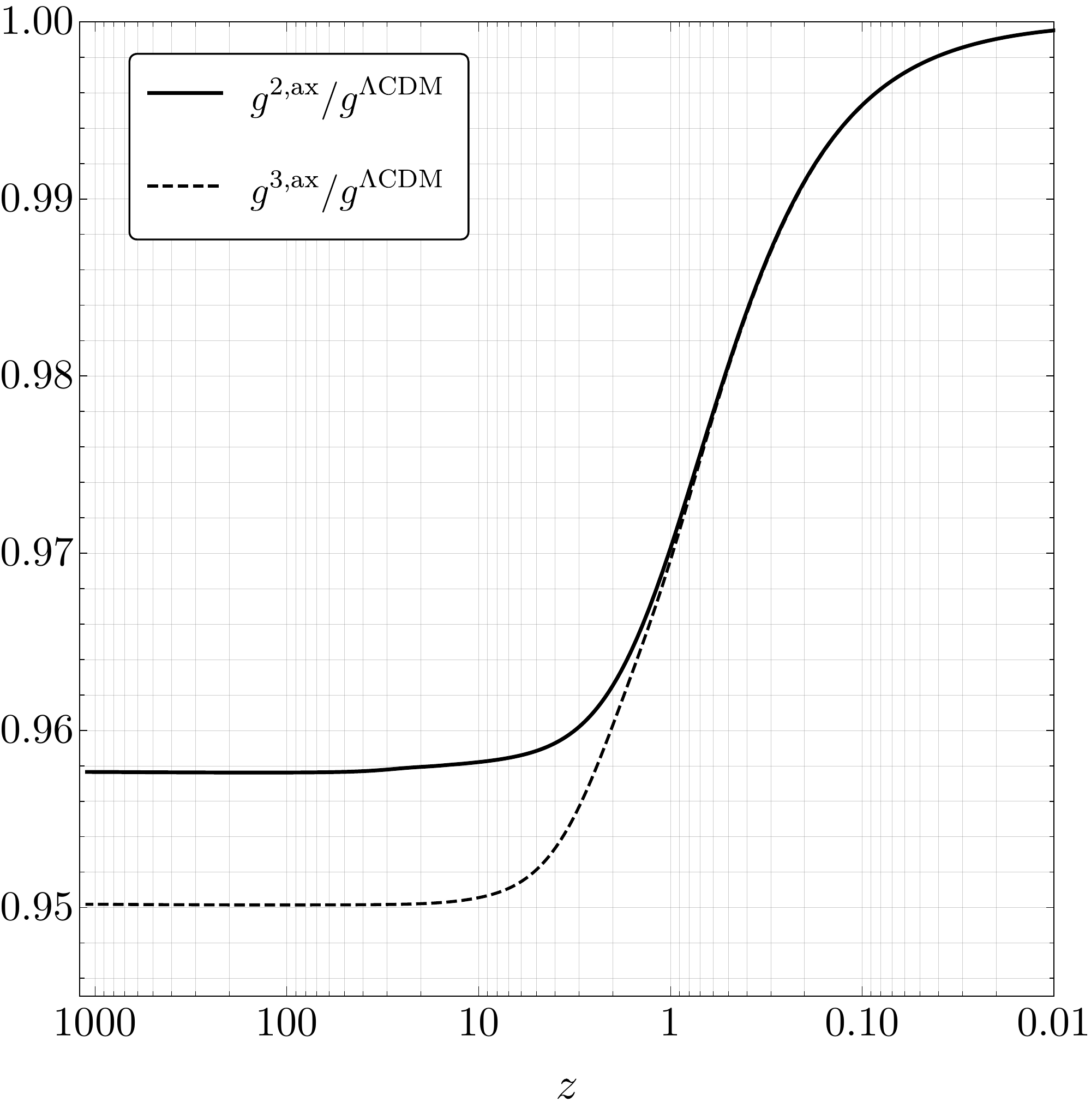}
	\centering
	\caption{Ratio of the growth functions of the two and three interacting ALP models to the growth function of the $\Lambda$CDM model at an order of $8 h^{-1} \, \text{Mpc}$ with parameters given in Table \ref{tab:ALP_params}.
	\label{fig:dm}}
\end{figure}

It is important to test these models against the $\sigma_8$ tension, which is milder than the $H_0$ tension, between the CMB and large scale structure (LSS) observations of the r.m.s linear fluctuation in the mass distribution on the scale of $8h^{-1}$Mpc. The following literature provides a framework for alleviating this tension \cite{DiValentino:2019dzu,DiValentino:2019ffd,Kumar:2019wfs,Pourtsidou:2016ico,An:2017crg,DiValentino:2018gcu,Kazantzidis:2018rnb}. We begin by solving the perturbation equations above numerically for the matter perturbations in these models. We then use the growth function formulation for quintessence models as described by \cite{Barros:2018efl}. 
We first define the density contrast to be,
\begin{align}
\delta(z) \equiv \frac{\delta\rho(z)}{\rho(z)}.
\end{align}
We then define the growth function which describes the development of the matter perturbations $g(z)$,
\begin{align}\label{eq:growth}
g(z) \equiv \frac{\delta(z)}{\delta_0},
\end{align}
where $\delta_0$ is the density contrast at $z=0$.
The models discussed here allow for parameters that increase the dark energy density compared to $\Lambda$CDM and decrease the matter energy density at late times as shown in the right panel of Fig. \ref{fig:ratio_dens}. In Fig. \ref{fig:dm}, we see that interactions slow down the development of the matter fluctuations which has an effect on the matter perturbations at late times. This means that galaxies will cluster slower compared to $\Lambda$CDM.

We test these models using KiDS-1000 \cite{Heymans:2020gsg}, which uses tomographic weak gravitational lensing. KiDS-1000 observes $S_{8,\Lambda} \equiv \sigma_8^{0}\sqrt{\frac{\Omega_m^{0}}{0.3}} = 0.766_{-0.014}^{+0.020}$, where $\sigma_8(0) \equiv \sigma_8^{0}$, using a $\Lambda$CDM cosmology. We also test these models against constraints found in the Dark Energy Survey (DES) \cite{Abbott:2021bzy}, which combines an analysis of galaxy clustering and weak gravitational lensing to obtain an $S_{8,\Lambda} = 0.776_{-0.017}^{+0.017}$.  Assuming a $\Lambda$CDM cosmology, the $S_8$ value at a given redshift $\bar{z}$ is,
\begin{align}\label{eq:S8_LCDM}
S_{8,\Lambda}(\bar{z}) = S_{8,\Lambda } g_{\Lambda }(\bar{z})\sqrt{\frac{\Omega_{m, \Lambda }(\bar{z})}{\Omega_{m, \Lambda}^0}},
\end{align}
where the `$\Lambda$' subscript refers to $\Lambda$CDM quantities and $g$ is the growth function given by Eq. (\ref{eq:growth}). To compare the models with data, we rescale Eq. (\ref{eq:S8_LCDM}) to obtain $S_8$ following \cite{Barros:2018efl},
\begin{align}
S_{8, \text{ax}} = S_{8, \Lambda}\frac{g_{\Lambda }(\bar{z})}{g_\text{ax}(\bar{z})}\sqrt{\frac{\Omega_{m,\Lambda }(\bar{z})}{\Omega_{m,\text{ax}}(\bar{z})}}\sqrt{\frac{\Omega_\text{ax}^0}{\Omega_{m, \Lambda }^{0}}}.
\end{align}
 KiDS-1000 and DES group galaxies together depending on their phenotypes in redshift bins from $z \approx 0.1$ to $z \approx 1$. Here, we take an approximately average value $\bar{z} = 0.5$ and express $\sigma_8^0$ as
\begin{align}
\label{eq: sigma8}
\sigma_{8,\text{ax}}^{0} = S_{8, \Lambda}\frac{g_{\Lambda }(\bar{z})}{g_\text{ax}(\bar{z})}\sqrt{\frac{\Omega_{m,\Lambda }(\bar{z})}{\Omega_{m,\text{ax}}(\bar{z})}}\sqrt{\frac{0.3}{\Omega_{m, \Lambda }^{0}}}.
\end{align}

	\begin{table}[H]
		\caption{The $\sigma_8^{0}$ values found using a $\Lambda$CDM model compared to the models of two and three interacting ALPs.}
		
		\begin{ruledtabular}
			\begin{tabular}{ccc}
				Model
				& KiDS-1000
				& DES
				\\ \hline
				$\Lambda$CDM\tablefootnote{The $\sigma_8^0$ values using the $\Lambda$CDM framework were taken from KiDS-1000 \cite{Heymans:2020gsg} and DES \cite{Abbott:2021bzy} data.}
				& $0.760^{+0.025}_{-0.020}$
				& $0.733^{+0.039}_{-0.049}$
				\\ 
				Two Interacting ALPs\footnotemark[2] 
				& $0.780_{-0.014}^{+0.020}$ 
				& $0.790_{-0.017}^{+0.017}$
				\\ 
				Three Interacting ALPs\footnotemark[2] 
				& $0.780_{-0.014}^{+0.020}$ 
				& $0.790_{-0.017}^{+0.017}$
				\\ 
			\end{tabular}
		\end{ruledtabular}
		\footnotetext[1]{The $\sigma_8^0$ values using the $\Lambda$CDM framework were taken from KiDS-1000 \cite{Heymans:2020gsg} and DES \cite{Abbott:2021bzy} data}
		\label{tab:ALP_results}
	\end{table}
\footnotetext[2]{ For the ALP models, we use an input value for $S_{8, \Lambda}$ equal to that measured by KiDS-1000 (second column) or DES (third column).}
We can compute the expected value of $\sigma_8^0$ in the models discussed here for either KiDS-1000 or DES by inputing the values for $S_{8,\Lambda}$ quoted above.  We can then compare these values to those found by Planck $\sigma_8^0 = 0.811^{+0.006}_{-0.006}$ \cite{2020}. One can see from Table \ref{tab:ALP_results} that the parameters in Table \ref{tab:ALP_params} chosen to alleviate the Hubble tension can also alleviate the $\sigma_8$ tension. 

\section{The Sound Horizon Problem}
\label{Sec: sound horizon}
The sound horizon at the epoch of baryon decoupling, $r_d$, is defined as the comoving distance a sound wave could travel from the beginning of the universe to the end of  matter-radiation decoupling. This quantity can be measured indirectly using observations from the CMB \cite{2020} and directly using BAO data \cite{Pogosian:2020ded,eBOSS:2020yzd}. The sound horizon problem arises when solutions to the $H_0$ tension introduce a further disagreement with BAO and CMB data. BAO data constrains the product of $Hr_d$ and favors a lower value of $r_d$ to agree with SH0ES.  However, Planck data favors higher values of $r_d$ \cite{DiValentino:2021izs,Bernal:2016gxb}. Hence, one way to agree with all data sets is to simultaneously increase $H_0$ and decrease the sound horizon. Late-time solutions, such as the models we describe in this work, run into tension with BAO data because they leave the sound horizon quantity unchanged. Although late-time solutions are slightly disfavoured \cite{Mortsell:2018mfj}, there is not enough evidence to rule them out completely. Early-time solutions that change the physics before recombination have been shown to alleviate the $H_0$ tension while satisfying BAO measurements by reducing $r_d$, but they can still run into tension with SH0ES \cite{Arendse:2019hev}. Furthermore, Krishnan et al. \cite{Krishnan:2020obg} investigated a multitude of late universe data and found that $H_0$ decreases with redshift, which gives more evidence to a similar trend proposed by strongly-lensed quasar time delay data (H0LiCOW) \cite{Wong:2019kwg}. This evidence potentially favours late-time solutions. However, it is worth noting that the trend is at a level of 2.1$\sigma$ and therefore does not give enough evidence against the continuance of early-time solutions. It has also been suggested that early-time solutions that reduce the sound horizon alone cannot completely resolve the $H_0$ tension because they run into tension with galaxy weak lensing data \cite{Jedamzik:2020zmd}. Lastly, it has been proposed that solutions to the $H_0$ tension must break one of seven assumptions \cite{Beenakker:2021vff}. The first four of these assumptions involve well-understood physics, which most models steer clear from breaking\footnotemark[3], while the last three are less fundamental and are the most common assumptions broken by early- and late-time solutions. Late-time solutions that do not change the sound horizon challenge the fifth assumption, which states that the estimation of the sound horizon due to Planck is accurate. \footnotetext[3]{The first four assumptions include: the physics of general relativity being an accurate estimation of the universe at cosmological scales, the universe can be approximated as spatially homogenous and isotropic, the universe has no spatial curvature, and the relationship between the photon redshift $z$ and scale factor $a$ is $a = \frac{1}{1+z}$\cite{Beenakker:2021vff}.} 

We now explore the effect the parameters in Table \ref{tab:ALP_params} have on the product $r_dh$, where $h \equiv H_0/(100 \, \text{km/s/Mpc})$ and $r_d = \int_{z_d}^{\infty}c_s(z)dz/H(z)$. We approximate the redshift of matter-radiation decoupling to be $z_d \approx 1100$ and $c_s(z)$ is the speed of sound of the photon-baryon fluid. In our model, we do not change the physics before matter-radiation decoupling, hence the sound horizon is unchanged. We may compare $(r_d)h_{\text{ax}}$ to $(r_dh)_{\text{CMB}}$ and  $(r_dh)_{\text{BAO}}$ by inputting the $r_d$ values measured by BAO and CMB, and then inputting $h_{\text{ax}}  \approx 0.74$ given in our model. BAO data measures a value of the sound horizon to be $r_d = 143.7 \pm 2.7 \, \text{Mpc}$ \cite{Pogosian:2020ded} and CMB data gives $r_d =  147.06 \pm 0.29 \, \text{Mpc}$\cite{2020}.
\begin{align}\label{eq:hrd}
\begin{matrix}
(r_dh)_{\text{BAO}} = (99.95 \pm 1.20) \, \text{Mpc},
\\
\\
(r_dh)_{\text{CMB}} = (99.08 \pm 0.80) \, \text{Mpc},
\\
\\
(r_{d, \, \text{BAO}})h_{ax} \approx (103.2 \pm 2.0) \, \text{Mpc},
\\
\\
(r_{d, \, \text{CMB}})h_{ax} \approx (108.82 \pm 0.21) \, \text{Mpc}.
\end{matrix}
\end{align}
 As mentioned before, $r_d$  does not change because we do not change the physice before $z_d \approx 1100$. However, $h$ will change given our parameters. Therefore, as common with late-time solutions, our parameters that seemingly solve the $H_0$ tension come into conflict with the determination of the sound horizon from BAO and CMB data. However, a full analysis against cosmological data sets is well motivated by  the following work \cite{Arendse:2019hev,Jedamzik:2020zmd,Beenakker:2021vff,Krishnan:2020obg}. 

\section{Conclusion}\label{sec:conclusion}
We analyze models of two and three interacting axion-like particles that act as dynamical dark energy.  As a proof of concept, we show that there exist parameters that may be controlled within both interacting frameworks that can alleviate both the Hubble tension and the less significant $\sigma_8$ tension simultaneously.  However, with our parameter values, these models comes into tension with BAO and Pantheon data. Hence, a full data analysis of the parameter spaces, considering a combination of constraints from cosmological data sets is to be pursued in future work.

\section{Acknowledgements}\label{sec:acknowledgements}
We thank Guido D'Amico for sharing his Python code for their interacting ALP model. We also thank Colin Bischoff, Joshua Eby, Cenalo Vaz, and Bin Wang for discussions.  E. M. and L. S. thank the Department of Physics at the University of Cincinnati for financial support in the form of the Joiner Fellowship and the Violet M. Diller Fellowship.  L. S. and L. C. R. W. thank the University of Cincinnati Office of Research Faculty Bridge Program for funding through the Faculty Bridge Grant.

\bibliography{dark_energy}

\providecommand{\href}[2]{#2}\begingroup\raggedright\begin{thebibliography}{10}

\bibitem{Condon:2018eqx}
J.~J. Condon and A.~M. Matthews, \emph{{$\Lambda$CDM Cosmology for
  Astronomers}}, \href{https://doi.org/10.1088/1538-3873/aac1b2}{\emph{Publ.
  Astron. Soc. Pac.} {\bfseries 130} (2018) 073001}
  [\href{https://arxiv.org/abs/1804.10047}{{\ttfamily 1804.10047}}].

\bibitem{2020}
N.~Aghanim, Y.~Akrami, M.~Ashdown, J.~Aumont, C.~Baccigalupi, M.~Ballardini
  et~al., \emph{Planck 2018 results},
  \href{https://doi.org/10.1051/0004-6361/201833910}{\emph{Astronomy \&
  Astrophysics} {\bfseries 641} (2020) A6}.

\bibitem{Riess:2020fzl}
A.~G. Riess, S.~Casertano, W.~Yuan, J.~B. Bowers, L.~Macri, J.~C. Zinn et~al.,
  \emph{{Cosmic Distances Calibrated to 1\% Precision with Gaia EDR3 Parallaxes
  and Hubble Space Telescope Photometry of 75 Milky Way Cepheids Confirm
  Tension with $\Lambda$CDM}},
  \href{https://doi.org/10.3847/2041-8213/abdbaf}{\emph{Astrophys. J. Lett.}
  {\bfseries 908} (2021) L6}
  [\href{https://arxiv.org/abs/2012.08534}{{\ttfamily 2012.08534}}].

\bibitem{Poulin:2018cxd}
V.~Poulin, T.~L. Smith, T.~Karwal and M.~Kamionkowski, \emph{{Early Dark Energy
  Can Resolve The Hubble Tension}},
  \href{https://doi.org/10.1103/PhysRevLett.122.221301}{\emph{Phys. Rev. Lett.}
  {\bfseries 122} (2019) 221301}
  [\href{https://arxiv.org/abs/1811.04083}{{\ttfamily 1811.04083}}].

\bibitem{Lin:2019qug}
M.-X. Lin, G.~Benevento, W.~Hu and M.~Raveri, \emph{{Acoustic Dark Energy:
  Potential Conversion of the Hubble Tension}},
  \href{https://doi.org/10.1103/PhysRevD.100.063542}{\emph{Phys. Rev. D}
  {\bfseries 100} (2019) 063542}
  [\href{https://arxiv.org/abs/1905.12618}{{\ttfamily 1905.12618}}].

\bibitem{Hill:2020osr}
J.~C. Hill, E.~McDonough, M.~W. Toomey and S.~Alexander, \emph{{Early dark
  energy does not restore cosmological concordance}},
  \href{https://doi.org/10.1103/PhysRevD.102.043507}{\emph{Phys. Rev. D}
  {\bfseries 102} (2020) 043507}
  [\href{https://arxiv.org/abs/2003.07355}{{\ttfamily 2003.07355}}].

\bibitem{Sakstein:2019fmf}
J.~Sakstein and M.~Trodden, \emph{{Early Dark Energy from Massive Neutrinos as
  a Natural Resolution of the Hubble Tension}},
  \href{https://doi.org/10.1103/PhysRevLett.124.161301}{\emph{Phys. Rev. Lett.}
  {\bfseries 124} (2020) 161301}
  [\href{https://arxiv.org/abs/1911.11760}{{\ttfamily 1911.11760}}].

\bibitem{Niedermann:2020dwg}
F.~Niedermann and M.~S. Sloth, \emph{{Resolving the Hubble tension with new
  early dark energy}},
  \href{https://doi.org/10.1103/PhysRevD.102.063527}{\emph{Phys. Rev. D}
  {\bfseries 102} (2020) 063527}
  [\href{https://arxiv.org/abs/2006.06686}{{\ttfamily 2006.06686}}].

\bibitem{DiValentino:2017iww}
E.~Di~Valentino, A.~Melchiorri and O.~Mena, \emph{{Can interacting dark energy
  solve the $H_0$ tension?}},
  \href{https://doi.org/10.1103/PhysRevD.96.043503}{\emph{Phys. Rev. D}
  {\bfseries 96} (2017) 043503}
  [\href{https://arxiv.org/abs/1704.08342}{{\ttfamily 1704.08342}}].

\bibitem{Yang:2018uae}
W.~Yang, A.~Mukherjee, E.~Di~Valentino and S.~Pan, \emph{{Interacting dark
  energy with time varying equation of state and the $H_0$ tension}},
  \href{https://doi.org/10.1103/PhysRevD.98.123527}{\emph{Phys. Rev. D}
  {\bfseries 98} (2018) 123527}
  [\href{https://arxiv.org/abs/1809.06883}{{\ttfamily 1809.06883}}].

\bibitem{DiValentino:2019ffd}
E.~Di~Valentino, A.~Melchiorri, O.~Mena and S.~Vagnozzi, \emph{{Interacting
  dark energy in the early 2020s: A promising solution to the $H_0$ and cosmic
  shear tensions}},
  \href{https://doi.org/10.1016/j.dark.2020.100666}{\emph{Phys. Dark Univ.}
  {\bfseries 30} (2020) 100666}
  [\href{https://arxiv.org/abs/1908.04281}{{\ttfamily 1908.04281}}].

\bibitem{Yang:2018euj}
W.~Yang, S.~Pan, E.~Di~Valentino, R.~C. Nunes, S.~Vagnozzi and D.~F. Mota,
  \emph{{Tale of stable interacting dark energy, observational signatures, and
  the $H_0$ tension}},
  \href{https://doi.org/10.1088/1475-7516/2018/09/019}{\emph{JCAP} {\bfseries
  09} (2018) 019} [\href{https://arxiv.org/abs/1805.08252}{{\ttfamily
  1805.08252}}].

\bibitem{Yang:2019uzo}
W.~Yang, O.~Mena, S.~Pan and E.~Di~Valentino, \emph{{Dark sectors with
  dynamical coupling}},
  \href{https://doi.org/10.1103/PhysRevD.100.083509}{\emph{Phys. Rev. D}
  {\bfseries 100} (2019) 083509}
  [\href{https://arxiv.org/abs/1906.11697}{{\ttfamily 1906.11697}}].

\bibitem{Wang:2016lxa}
B.~Wang, E.~Abdalla, F.~Atrio-Barandela and D.~Pavon, \emph{{Dark Matter and
  Dark Energy Interactions: Theoretical Challenges, Cosmological Implications
  and Observational Signatures}},
  \href{https://doi.org/10.1088/0034-4885/79/9/096901}{\emph{Rept. Prog. Phys.}
  {\bfseries 79} (2016) 096901}
  [\href{https://arxiv.org/abs/1603.08299}{{\ttfamily 1603.08299}}].

\bibitem{Aghanim:2018eyx}
{\scshape Planck} collaboration, \emph{{Planck 2018 results. VI. Cosmological
  parameters}},
  \href{https://doi.org/10.1051/0004-6361/201833910}{\emph{Astron. Astrophys.}
  {\bfseries 641} (2020) A6}
  [\href{https://arxiv.org/abs/1807.06209}{{\ttfamily 1807.06209}}].

\bibitem{DiValentino:2019dzu}
E.~Di~Valentino, A.~Melchiorri and J.~Silk, \emph{{Cosmological constraints in
  extended parameter space from the Planck 2018 Legacy release}},
  \href{https://doi.org/10.1088/1475-7516/2020/01/013}{\emph{JCAP} {\bfseries
  01} (2020) 013} [\href{https://arxiv.org/abs/1908.01391}{{\ttfamily
  1908.01391}}].

\bibitem{Yang:2020zuk}
W.~Yang, E.~Di~Valentino, S.~Pan, S.~Basilakos and A.~Paliathanasis,
  \emph{{Metastable dark energy models in light of $Planck$ 2018 data:
  Alleviating the $H_0$ tension}},
  \href{https://doi.org/10.1103/PhysRevD.102.063503}{\emph{Phys. Rev. D}
  {\bfseries 102} (2020) 063503}
  [\href{https://arxiv.org/abs/2001.04307}{{\ttfamily 2001.04307}}].

\bibitem{Cai:2021wgv}
R.-G. Cai, Z.-K. Guo, L.~Li, S.-J. Wang and W.-W. Yu, \emph{{Chameleon dark
  energy can resolve the Hubble tension}},
  \href{https://arxiv.org/abs/2102.02020}{{\ttfamily 2102.02020}}.

\bibitem{Wang:2016och}
S.~Wang, Y.~Wang and M.~Li, \emph{{Holographic Dark Energy}},
  \href{https://doi.org/10.1016/j.physrep.2017.06.003}{\emph{Phys. Rept.}
  {\bfseries 696} (2017) 1} [\href{https://arxiv.org/abs/1612.00345}{{\ttfamily
  1612.00345}}].

\bibitem{DiValentino:2021izs}
E.~Di~Valentino, O.~Mena, S.~Pan, L.~Visinelli, W.~Yang, A.~Melchiorri et~al.,
  \emph{{In the Realm of the Hubble tension $-$ a Review of Solutions}},
  \href{https://arxiv.org/abs/2103.01183}{{\ttfamily 2103.01183}}.

\bibitem{Knox:2019rjx}
L.~Knox and M.~Millea, \emph{{Hubble constant hunter\textquoteright{}s guide}},
  \href{https://doi.org/10.1103/PhysRevD.101.043533}{\emph{Phys. Rev. D}
  {\bfseries 101} (2020) 043533}
  [\href{https://arxiv.org/abs/1908.03663}{{\ttfamily 1908.03663}}].

\bibitem{Arendse:2019hev}
N.~Arendse et~al., \emph{{Cosmic dissonance: are new physics or systematics
  behind a short sound horizon?}},
  \href{https://doi.org/10.1051/0004-6361/201936720}{\emph{Astron. Astrophys.}
  {\bfseries 639} (2020) A57}
  [\href{https://arxiv.org/abs/1909.07986}{{\ttfamily 1909.07986}}].

\bibitem{Jedamzik:2020zmd}
K.~Jedamzik, L.~Pogosian and G.-B. Zhao, \emph{{Why reducing the cosmic sound
  horizon can not fully resolve the Hubble tension}},
  \href{https://arxiv.org/abs/2010.04158}{{\ttfamily 2010.04158}}.

\bibitem{Beenakker:2021vff}
W.~Beenakker and D.~Venhoek, \emph{{A structured analysis of Hubble tension}},
  \href{https://arxiv.org/abs/2101.01372}{{\ttfamily 2101.01372}}.

\bibitem{Krishnan:2020obg}
C.~Krishnan, E.~O. Colg\'ain, Ruchika, A.~A. Sen, M.~M. Sheikh-Jabbari and
  T.~Yang, \emph{{Is there an early Universe solution to Hubble tension?}},
  \href{https://doi.org/10.1103/PhysRevD.102.103525}{\emph{Phys. Rev. D}
  {\bfseries 102} (2020) 103525}
  [\href{https://arxiv.org/abs/2002.06044}{{\ttfamily 2002.06044}}].

\bibitem{DAmico:2016jbm}
G.~D'Amico, T.~Hamill and N.~Kaloper, \emph{{Quantum field theory of
  interacting dark matter and dark energy: Dark monodromies}},
  \href{https://doi.org/10.1103/PhysRevD.94.103526}{\emph{Phys. Rev. D}
  {\bfseries 94} (2016) 103526}
  [\href{https://arxiv.org/abs/1605.00996}{{\ttfamily 1605.00996}}].

\bibitem{Arias:2012az}
P.~Arias, D.~Cadamuro, M.~Goodsell, J.~Jaeckel, J.~Redondo and A.~Ringwald,
  \emph{{WISPy Cold Dark Matter}},
  \href{https://doi.org/10.1088/1475-7516/2012/06/013}{\emph{JCAP} {\bfseries
  06} (2012) 013} [\href{https://arxiv.org/abs/1201.5902}{{\ttfamily
  1201.5902}}].

\bibitem{Visinelli:2017imh}
L.~Visinelli, \emph{{Light axion-like dark matter must be present during
  inflation}}, \href{https://doi.org/10.1103/PhysRevD.96.023013}{\emph{Phys.
  Rev. D} {\bfseries 96} (2017) 023013}
  [\href{https://arxiv.org/abs/1703.08798}{{\ttfamily 1703.08798}}].

\bibitem{Hlozek:2014lca}
R.~Hlozek, D.~Grin, D.~J.~E. Marsh and P.~G. Ferreira, \emph{{A search for
  ultralight axions using precision cosmological data}},
  \href{https://doi.org/10.1103/PhysRevD.91.103512}{\emph{Phys. Rev. D}
  {\bfseries 91} (2015) 103512}
  [\href{https://arxiv.org/abs/1410.2896}{{\ttfamily 1410.2896}}].

\bibitem{Visinelli:2018utg}
L.~Visinelli and S.~Vagnozzi, \emph{{Cosmological window onto the string
  axiverse and the supersymmetry breaking scale}},
  \href{https://doi.org/10.1103/PhysRevD.99.063517}{\emph{Phys. Rev. D}
  {\bfseries 99} (2019) 063517}
  [\href{https://arxiv.org/abs/1809.06382}{{\ttfamily 1809.06382}}].

\bibitem{Fung:2021fcj}
L.~W. Fung, L.~Li, T.~Liu, H.~N. Luu, Y.-C. Qiu and S.~H.~H. Tye, \emph{{The
  Hubble Constant in the Axi-Higgs Universe}},
  \href{https://arxiv.org/abs/2105.01631}{{\ttfamily 2105.01631}}.

\bibitem{Fung:2021wbz}
L.~W. Fung, L.~Li, T.~Liu, H.~N. Luu, Y.-C. Qiu and S.~H.~H. Tye,
  \emph{{Axi-Higgs Cosmology}},
  \href{https://arxiv.org/abs/2102.11257}{{\ttfamily 2102.11257}}.

\bibitem{Riess_2011}
A.~G. Riess, L.~Macri, S.~Casertano, H.~Lampeitl, H.~C. Ferguson, A.~V.
  Filippenko et~al., \emph{A 3
  with thehubble space telescopeand wide field camera 3},
  \href{https://doi.org/10.1088/0004-637x/730/2/119}{\emph{The Astrophysical
  Journal} {\bfseries 730} (2011) 119}.

\bibitem{Riess_2016}
A.~G. Riess, L.~M. Macri, S.~L. Hoffmann, D.~Scolnic, S.~Casertano, A.~V.
  Filippenko et~al., \emph{A 2.4
  hubble constant}, \href{https://doi.org/10.3847/0004-637x/826/1/56}{\emph{The
  Astrophysical Journal} {\bfseries 826} (2016) 56}.

\bibitem{Riess_2019}
A.~G. Riess, S.~Casertano, W.~Yuan, L.~M. Macri and D.~Scolnic, \emph{Large
  magellanic cloud cepheid standards provide a 1
  determination of the hubble constant and stronger evidence for physics beyond
  $\lambda$cdm}, \href{https://doi.org/10.3847/1538-4357/ab1422}{\emph{The
  Astrophysical Journal} {\bfseries 876} (2019) 85}.

\bibitem{Macaulay:2013swa}
E.~Macaulay, I.~K. Wehus and H.~K. Eriksen, \emph{{Lower Growth Rate from
  Recent Redshift Space Distortion Measurements than Expected from Planck}},
  \href{https://doi.org/10.1103/PhysRevLett.111.161301}{\emph{Phys. Rev. Lett.}
  {\bfseries 111} (2013) 161301}
  [\href{https://arxiv.org/abs/1303.6583}{{\ttfamily 1303.6583}}].

\bibitem{Ade:2015xua}
{\scshape Planck} collaboration, \emph{{Planck 2015 results. XIII. Cosmological
  parameters}},
  \href{https://doi.org/10.1051/0004-6361/201525830}{\emph{Astron. Astrophys.}
  {\bfseries 594} (2016) A13}
  [\href{https://arxiv.org/abs/1502.01589}{{\ttfamily 1502.01589}}].

\bibitem{Heymans:2020gsg}
C.~Heymans et~al., \emph{{KiDS-1000 Cosmology: Multi-probe weak gravitational
  lensing and spectroscopic galaxy clustering constraints}},
  \href{https://doi.org/10.1051/0004-6361/202039063}{\emph{Astron. Astrophys.}
  {\bfseries 646} (2021) A140}
  [\href{https://arxiv.org/abs/2007.15632}{{\ttfamily 2007.15632}}].

\bibitem{Abbott:2017wau}
{\scshape DES} collaboration, \emph{{Dark Energy Survey year 1 results:
  Cosmological constraints from galaxy clustering and weak lensing}},
  \href{https://doi.org/10.1103/PhysRevD.98.043526}{\emph{Phys. Rev. D}
  {\bfseries 98} (2018) 043526}
  [\href{https://arxiv.org/abs/1708.01530}{{\ttfamily 1708.01530}}].

\bibitem{Barros:2018efl}
B.~J. Barros, L.~Amendola, T.~Barreiro and N.~J. Nunes, \emph{{Coupled
  quintessence with a $\Lambda$CDM background: removing the $\sigma_8$
  tension}}, \href{https://doi.org/10.1088/1475-7516/2019/01/007}{\emph{JCAP}
  {\bfseries 01} (2019) 007}
  [\href{https://arxiv.org/abs/1802.09216}{{\ttfamily 1802.09216}}].

\bibitem{Marsh:2015xka}
D.~J.~E. Marsh, \emph{{Axion Cosmology}},
  \href{https://doi.org/10.1016/j.physrep.2016.06.005}{\emph{Phys. Rept.}
  {\bfseries 643} (2016) 1} [\href{https://arxiv.org/abs/1510.07633}{{\ttfamily
  1510.07633}}].

\bibitem{Hildebrandt:2016iqg}
H.~Hildebrandt et~al., \emph{{KiDS-450: Cosmological parameter constraints from
  tomographic weak gravitational lensing}},
  \href{https://doi.org/10.1093/mnras/stw2805}{\emph{Mon. Not. Roy. Astron.
  Soc.} {\bfseries 465} (2017) 1454}
  [\href{https://arxiv.org/abs/1606.05338}{{\ttfamily 1606.05338}}].

\bibitem{Lewis:1999bs}
A.~Lewis, A.~Challinor and A.~Lasenby, \emph{{Efficient computation of CMB
  anisotropies in closed FRW models}},
  \href{https://doi.org/10.1086/309179}{\emph{Astrophys. J.} {\bfseries 538}
  (2000) 473} [\href{https://arxiv.org/abs/astro-ph/9911177}{{\ttfamily
  astro-ph/9911177}}].

\bibitem{Kumar:2019wfs}
S.~Kumar, R.~C. Nunes and S.~K. Yadav, \emph{{Dark sector interaction: a remedy
  of the tensions between CMB and LSS data}},
  \href{https://doi.org/10.1140/epjc/s10052-019-7087-7}{\emph{Eur. Phys. J. C}
  {\bfseries 79} (2019) 576}
  [\href{https://arxiv.org/abs/1903.04865}{{\ttfamily 1903.04865}}].

\bibitem{Pourtsidou:2016ico}
A.~Pourtsidou and T.~Tram, \emph{{Reconciling CMB and structure growth
  measurements with dark energy interactions}},
  \href{https://doi.org/10.1103/PhysRevD.94.043518}{\emph{Phys. Rev. D}
  {\bfseries 94} (2016) 043518}
  [\href{https://arxiv.org/abs/1604.04222}{{\ttfamily 1604.04222}}].

\bibitem{An:2017crg}
R.~An, C.~Feng and B.~Wang, \emph{{Relieving the Tension between Weak Lensing
  and Cosmic Microwave Background with Interacting Dark Matter and Dark Energy
  Models}}, \href{https://doi.org/10.1088/1475-7516/2018/02/038}{\emph{JCAP}
  {\bfseries 02} (2018) 038}
  [\href{https://arxiv.org/abs/1711.06799}{{\ttfamily 1711.06799}}].

\bibitem{DiValentino:2018gcu}
E.~Di~Valentino and S.~Bridle, \emph{{Exploring the Tension between Current
  Cosmic Microwave Background and Cosmic Shear Data}},
  \href{https://doi.org/10.3390/sym10110585}{\emph{Symmetry} {\bfseries 10}
  (2018) 585}.

\bibitem{Kazantzidis:2018rnb}
L.~Kazantzidis and L.~Perivolaropoulos, \emph{{Evolution of the $f\sigma_8$
  tension with the Planck15/$\Lambda$CDM determination and implications for
  modified gravity theories}},
  \href{https://doi.org/10.1103/PhysRevD.97.103503}{\emph{Phys. Rev. D}
  {\bfseries 97} (2018) 103503}
  [\href{https://arxiv.org/abs/1803.01337}{{\ttfamily 1803.01337}}].

\bibitem{Abbott:2021bzy}
{\scshape DES} collaboration, \emph{{Dark Energy Survey Year 3 Results:
  Cosmological Constraints from Galaxy Clustering and Weak Lensing}},
  \href{https://arxiv.org/abs/2105.13549}{{\ttfamily 2105.13549}}.

\bibitem{Pogosian:2020ded}
L.~Pogosian, G.-B. Zhao and K.~Jedamzik, \emph{{Recombination-independent
  determination of the sound horizon and the Hubble constant from BAO}},
  \href{https://doi.org/10.3847/2041-8213/abc6a8}{\emph{Astrophys. J. Lett.}
  {\bfseries 904} (2020) L17}
  [\href{https://arxiv.org/abs/2009.08455}{{\ttfamily 2009.08455}}].

\bibitem{eBOSS:2020yzd}
{\scshape eBOSS} collaboration, \emph{{Completed SDSS-IV extended Baryon
  Oscillation Spectroscopic Survey: Cosmological implications from two decades
  of spectroscopic surveys at the Apache Point Observatory}},
  \href{https://doi.org/10.1103/PhysRevD.103.083533}{\emph{Phys. Rev. D}
  {\bfseries 103} (2021) 083533}
  [\href{https://arxiv.org/abs/2007.08991}{{\ttfamily 2007.08991}}].

\bibitem{Bernal:2016gxb}
J.~L. Bernal, L.~Verde and A.~G. Riess, \emph{{The trouble with $H_0$}},
  \href{https://doi.org/10.1088/1475-7516/2016/10/019}{\emph{JCAP} {\bfseries
  10} (2016) 019} [\href{https://arxiv.org/abs/1607.05617}{{\ttfamily
  1607.05617}}].

\bibitem{Mortsell:2018mfj}
E.~M\"ortsell and S.~Dhawan, \emph{{Does the Hubble constant tension call for
  new physics?}},
  \href{https://doi.org/10.1088/1475-7516/2018/09/025}{\emph{JCAP} {\bfseries
  09} (2018) 025} [\href{https://arxiv.org/abs/1801.07260}{{\ttfamily
  1801.07260}}].

\bibitem{Wong:2019kwg}
K.~C. Wong et~al., \emph{{H0LiCOW \textendash{} XIII. A 2.4 per cent
  measurement of H0 from lensed quasars: 5.3\ensuremath{\sigma} tension between
  early- and late-Universe probes}},
  \href{https://doi.org/10.1093/mnras/stz3094}{\emph{Mon. Not. Roy. Astron.
  Soc.} {\bfseries 498} (2020) 1420}
  [\href{https://arxiv.org/abs/1907.04869}{{\ttfamily 1907.04869}}].

\end{thebibliography}\endgroup
\bibliographystyle{JHEP}

\end{document}